\documentclass[english,aps,prapplied,twocolumn,superscriptaddress,longbibliography]{revtex4-2}
\usepackage[urlcolor=blue, hyperindex, colorlinks, bookmarks=true]{hyperref}

\usepackage{graphicx}
\usepackage{dcolumn}
\usepackage{bm}
\usepackage{amsmath,color,amssymb}
\usepackage[normalem]{ulem}
\usepackage{bm}
\usepackage{soul}
\usepackage[utf8]{inputenc}
\usepackage[T1]{fontenc}
\usepackage{mathptmx}
\usepackage{mathtools}
\usepackage{float}
\RequirePackage[normalem]{ulem} 

\definecolor{new}{rgb}{.38,.6,.38}
\definecolor{old}{rgb}{1,0,0}
\definecolor{off}{rgb}{0,0,0}

\newcommand{\UCLAPhysics}{\affiliation{Department of Physics and Astronomy, UCLA, Los Angeles, CA 90095, USA}}
\newcommand{\UCLACQSE}{\affiliation{Center for Quantum Science and Engineering, UCLA, Los Angeles, CA 90095, USA}}
\newcommand{\UCLAECE}{\affiliation{Department of Electrical and Computer Engineering, UCLA, Los Angeles, CA 90095, USA}}
\newcommand{\UCLAMath}{\affiliation{Department of Mathematics, UCLA, Los Angeles, CA 90095, USA}}
\newcommand{\PrincetonECE}{\affiliation{Department of Electrical and Computer Engineering, Princeton University, Princeton, NJ 08544, USA}}
\newcommand{\PrincetonPhysics}{\affiliation{Department of Physics, Princeton University, Princeton, NJ 08544, USA}}

\begin{document}
\title{Towards Utilizing Scanning Gate Microscopy as a High-Resolution Probe of Valley Splitting in Si/SiGe Heterostructures}

\author{Efe Cakar}
\UCLAPhysics
\UCLACQSE
\author{H. Ekmel Ercan}
\UCLAECE
\author{Gordian Fuchs}
\PrincetonECE
\author{Artem O. Denisov}
\PrincetonPhysics
\author{Christopher R. Anderson}
\UCLACQSE
\UCLAMath
\author{Mark F. Gyure}
\UCLACQSE
\UCLAECE
\author{Jason R. Petta}
\UCLAPhysics
\UCLACQSE

\begin{abstract}
A detailed understanding of the material properties that affect the splitting between the two low-lying valley states in Si/SiGe heterostructures will be increasingly important as the number of spin qubits is increased. Scanning gate microscopy has been proposed as a method to measure the spatial variation of the valley splitting as a tip-induced dot is moved around in the plane of the Si quantum well. We develop a simulation using an electrostatic model of the scanning gate microscope tip and the overlapping gate structure combined with an approximate solution to the three-dimensional Schrödinger-Poisson equation in the device stack. Using this simulation, we show that a tip-induced quantum dot formed near source and drain electrodes can be adiabatically moved to a region far from the gate electrodes. We argue that by spatially translating the tip-induced dot across a defect in the Si/SiGe interface, changes in valley splitting can be detected.
\end{abstract}

\maketitle

Semiconductor spin qubits have emerged as top contenders for quantum computation \cite{burkard2023semiconductor}, owing to their long coherence times \cite{tyryshkin2012electron}, high single- and two-qubit control fidelities \cite{mills2022two, xue2022quantum, noiri2022fast}, and scalability \cite{zajac2016scalable}. On the other hand, there are materials challenges in silicon quantum devices that may impede the ability to scale up to larger system sizes \cite{burkard2023semiconductor}. For example, charge noise is known to impact the performance of spin qubits, as the gate-voltage tunable exchange interaction opens up sensitivity to voltage fluctuations \cite{freeman2016comparison, mi2018landau, connors2022charge}. In addition, silicon possesses a six-fold valley degeneracy \cite{ando1982electronic, burkard2016dispersive}, which is partially lifted due to the Si/SiGe lattice mismatch \cite{schaffler1997high, Zwanenburg2013silicon}. The energy splitting of the two lowest-lying valley states, commonly referred to as the valley splitting, can be comparable to single spin qubit Zeeman energies \cite{zajac2015a, mi2017high, hollman2020large, chen2021detuning}. As these valley states can provide decoherence paths and negatively impact the control and readout of Si/SiGe spin qubits, there is demand for characterization approaches that have the potential to correlate material properties with variations in valley splitting and ultimately with qubit performance.

Consistently large valley splittings are desired as Si/SiGe quantum devices are scaled up. However, valley splittings can vary even in devices fabricated on the same chip, ranging from $\sim$20--300~$\mu$eV \cite{Borselli2011measurement, zajac2015a, mi2017high, Schoenfield2017coherent, hollman2020large, chen2021detuning}. Since atomistic \cite{friesen2006magnetic, kharche2007valley, culcer2010interface, gamble2013disorder, boross2016control, hosseinkhani2020electromagnetic, dodson2022how, EkmelPRB} and alloy \cite{Wuetz2022atomic, Losert2023practical, pena2023utilizing, volmer2023mapping, esposti2024low, EkmelarXiv} disorder have been shown to influence valley states, an efficient probe of valley splitting is desirable. Scanning gate microscopy (SGM) has been proposed as a method to measure the spatial variation of valley splitting \cite{shim2019induced} and may be used as a tool to characterize structural disorder in Si/SiGe heterostructures. The effects of small variations of donor positions on valley states have recently been shown for P donors in Si using scanning tunneling microscopy \cite{voison2020valley}. However, despite recent efforts in using SGM to control and read out quantum dots \cite{Denisov2022microwave, denisov2023dispersive}, the ability to probe valley splitting using an SGM tip has yet to be realized for buried heterostructures.

\begin{figure}[t]
	\centering
	\includegraphics[width=\columnwidth]{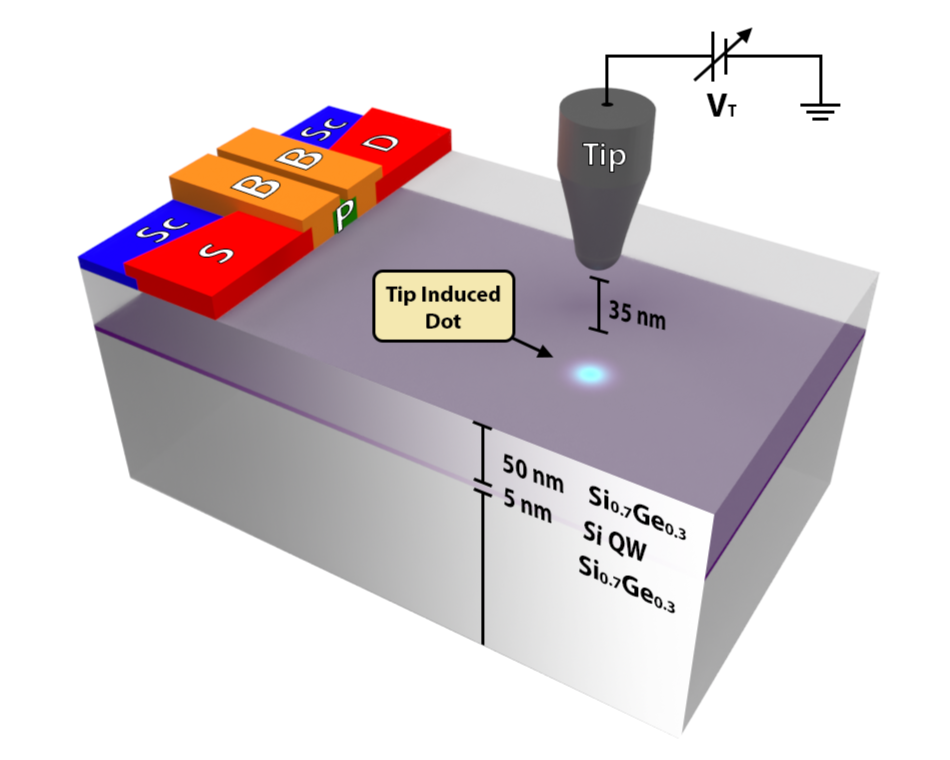}
	\caption{Schematic illustration of the proposed experimental setup. The Si/SiGe heterostructure consists of a 5 nm thick Si QW that is grown on top of a Si$_{0.7}$Ge$_{0.3}$ spacer layer and capped with a 50~nm thick Si$_{0.7}$Ge$_{0.3}$ upper spacer layer. The plunger, barrier, source, drain, and screening gate electrodes are controlled with voltage biases V$_{P}$, V$_{B}$, V$_{S}$, V$_{D}$, and V$_{Sc}$, respectively. The SGM tip is biased at a potential V$_{T}$. It can move freely 35 nm above the sample surface. The tip-induced dot is depicted in light blue.
}
	\label{fig:1}
\end{figure}

\begin{figure*}[t!]
	\centering
	\includegraphics[width=2\columnwidth]{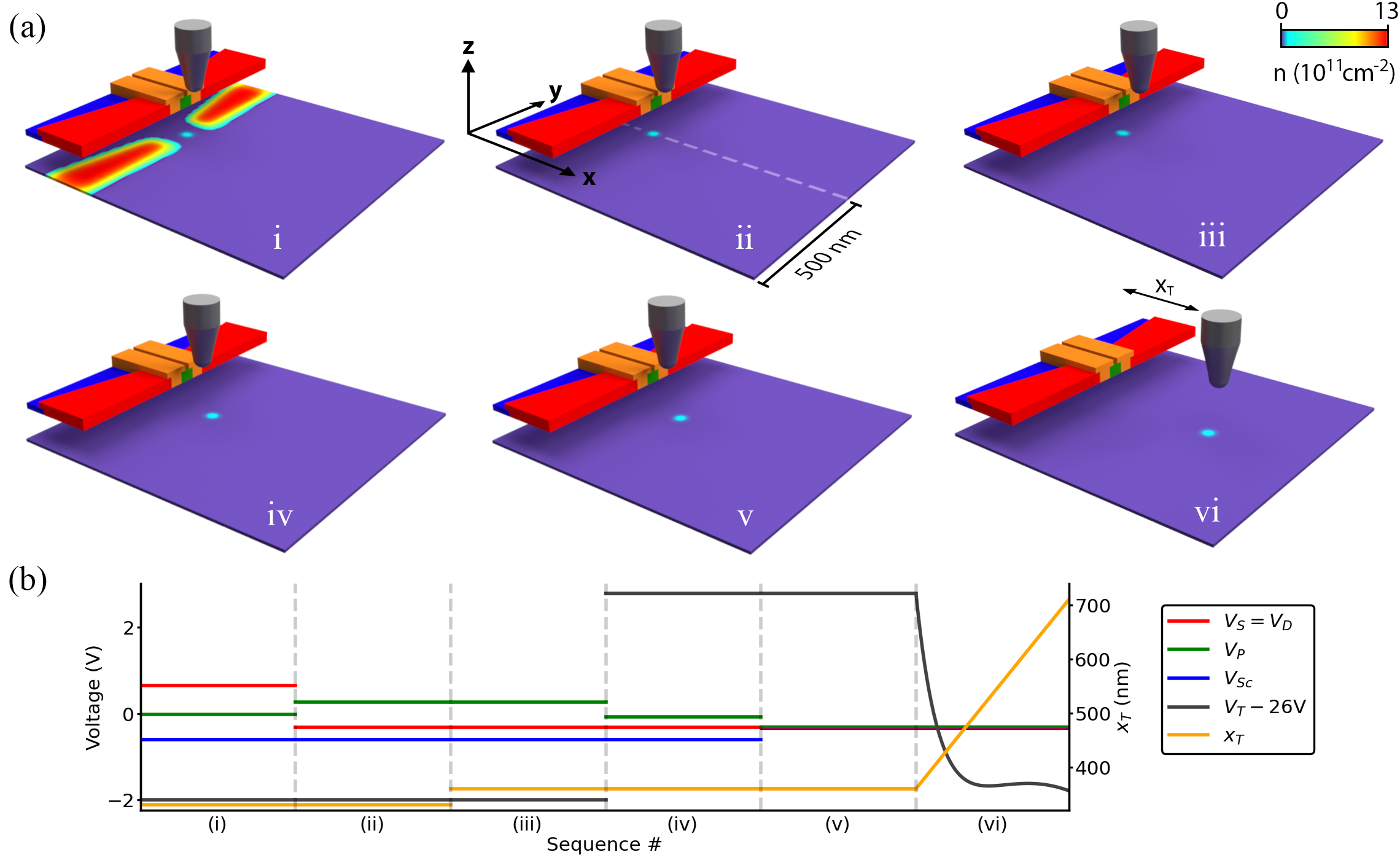}
	\caption{Electronic loading protocol. (a) Electron densities in the Si QW are shown for six different gate bias and tip position configurations as illustrated in (b). The loading sequence is as follows: (i) The reservoirs under the source and drain gates are populated and a single electron occupies the region under the plunger gate. (ii) Source and drain voltages are lowered to deplete the reservoirs and the plunger gate bias is increased to maintain confinement of the single electron. (iii) The tip is moved 30 nm in the $x$-direction. (iv) The plunger gate voltage is lowered as the tip voltage is increased to transfer a single electron from a QD formed under the patterned gate electrodes to the potential minimum created by the voltage-biased SGM tip. (v) Source, drain, plunger and screening biases are set equal to reduce variations in the potential. (vi) The tip is moved 350 nm in the $x$-direction and the tip bias is decreased to ensure the adiabatic translation of the single electron dot. (b) Gate voltages and tip position x$_{T}$ for steps (i) -- (vi). Note that V$_{T}$ is offset by -26 V on the plot for ease of comparison with the other gate voltages. The barrier gate voltage is held constant at V$_{B}$ = -0.8 V.}
	\label{fig:2}
\end{figure*}

In this Letter, we investigate the feasibility of using a movable SGM tip to probe the valley splitting in a Si/SiGe heterostructure. Based on previous work \cite{zajac2015a, Denisov2022microwave}, we propose a gate structure where a SGM tip is biased to induce a quantum dot (QD), which is adiabatically moved to a region far from the gate electrodes. The tip-induced QD is then used to probe a step edge in the Si/SiGe interface by detecting changes in the valley splitting. We utilize composite overlapping mesh discretization \cite{chesshire1990composite} and efficient boundary conditions to achieve fast computations of varying gate voltages and tip positions. These methods allow us to build on the electronic confinement and valley state simulations of similar devices \cite{anderson2022high, dodson2022how} by including the gate geometry as a variable parameter and resolving the variation in valley splitting with sufficient precision.

Figure 1 illustrates the proposed experimental setup. The objective is to use a SGM tip to induce the formation of a QD in the silicon quantum well (QW). The Si/SiGe heterostructure consists of a lower Si$_{0.7}$Ge$_{0.3}$ spacer layer, a 5 nm thick Si QW, and a 50 nm thick upper Si$_{0.7}$Ge$_{0.3}$ spacer layer \cite{zajac2015a}. A stacked gate electrode structure is used with screening gates in the first layer; source, drain, and plunger gates in the second layer; and barrier gates in the third layer \cite{angus2007gate, zajac2016scalable, dodson2020fabrication, philips2022universal, cai2023coherent}. The SGM tip can move freely across the $xy$-plane and is held 35~nm above the sample surface. The proposed geometry was chosen because early simulations indicated difficulties in obtaining single electron isolation under the tip by tunnelling directly from a reservoir underneath a simpler lone gate structure. As we will demonstrate for the proposed device, the tip can be biased at a voltage V$_{T}$ to induce a QD underneath and enable single electron occupation.

Device simulations performed to date have modeled conventional geometries where gate electrode stacks remain fixed on top of the Si/SiGe heterostructure \cite{anderson2022high, dodson2022how, Denisov2022microwave}. The electron density in the Si QW is then modeled as a function of gate bias voltages using commercial software, such as COMSOL \cite{zajac2015a} or custom Schrödinger-Poisson solvers \cite{anderson2009efficient}. Our approach uses the Schrödinger-Poisson method. However, finding solutions with the proposed geometry is challenging, due to the high grid resolution needed to account for the moving electron and the numerical noise associated with regridding.

To achieve the required precision and lower computation time, we used a multi-domain/multi-model simulation with a composite overlapping mesh discretization \cite{chesshire1990composite}. The mesh consisted of two components, a global grid for calculating the reservoir charge and a subdomain grid for calculating charge densities and electronic states of the QD. A single electron is always present in the quantum subdomain calculation, but the charge in the reservoir is variable. The gates are biased such that the charge distribution in the reservoir and the charge in the QD subdomain do not overlap. Additionally, an effective boundary condition operator, based on domain decomposition techniques \cite{keyes1995domain}, was used when determining the effects of gate biases on the QW potential. The coefficients of the effective boundary condition operator do not depend on the applied biases and only need to be determined once for a particular gate geometry. Moreover, the tip position is parameterized in the effective boundary condition operator, eliminating the need for regridding as the tip is moved. For a specific tip position and gate biases, the planar boundary values on the device stack are determined by solving a non-iterative matrix equation, resulting in a computational technique with low numerical noise.

\begin{figure}[tbp]
	\centering
	\includegraphics[width=\columnwidth]{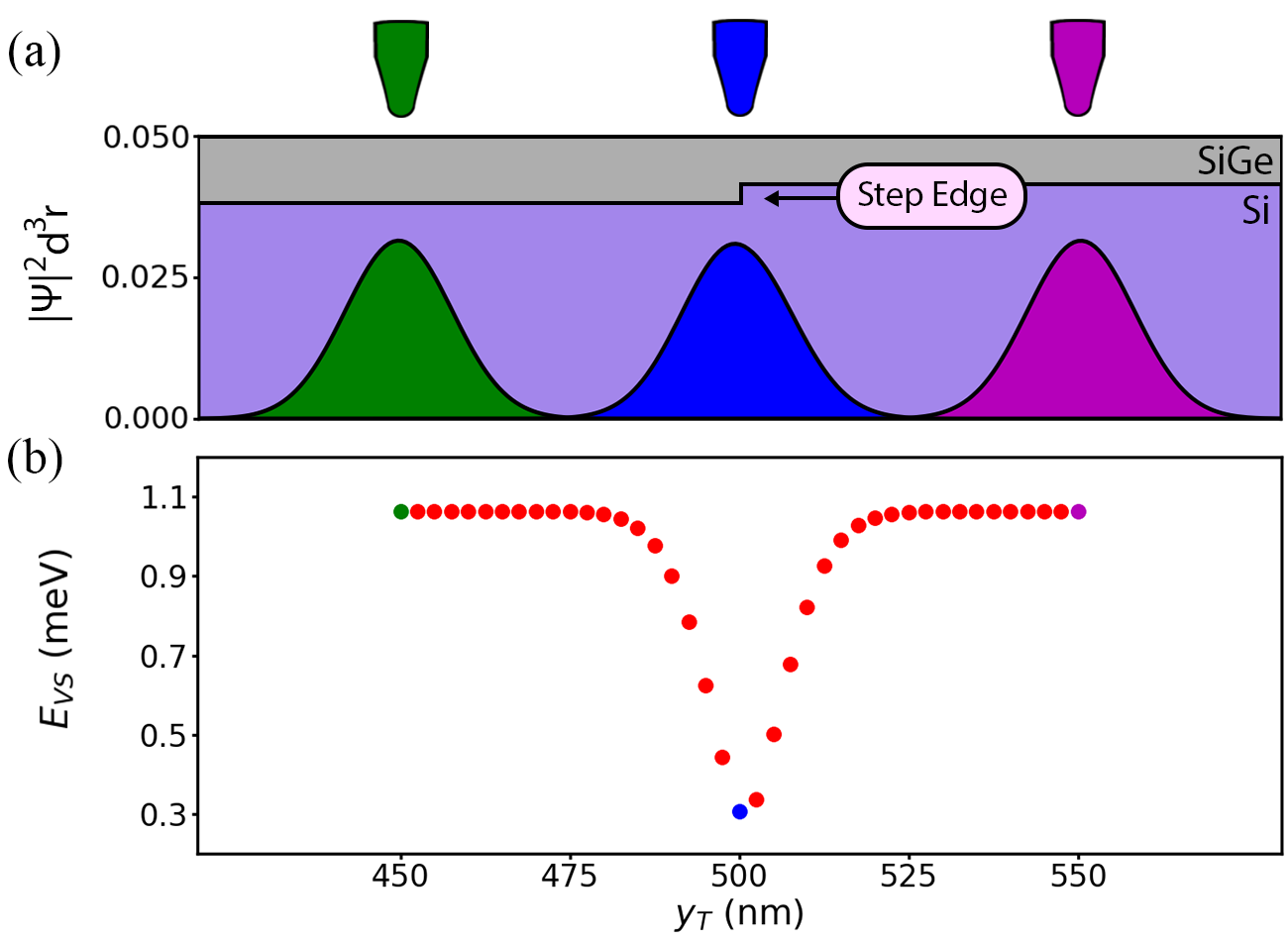}
	\caption{Variation of the valley splitting as the tip-induced dot is moved across the heterostructure. (a) A single atomic step in the upper Si/SiGe interface is located at $y$~=~500~nm. Electron probability amplitudes for tip positions, $y_{\rm T}$ = 450, 500, and 550~nm, are shown on the same plot. (b) Calculated valley splitting plotted as a function of $y_{\rm T}$. The green, blue and purple points correspond to the tip positions shown in panel (a).}
	\label{fig:3}
\end{figure}

Using this simulation, we were able to identify a protocol for dragging an electron across the heterostructure as illustrated in Fig.\ 2. Initially, the device is tuned such that the reservoirs under the source and drain gates have high electron densities and a single electron QD is formed under the plunger gate. Subsequently, the reservoirs are depleted by lowering the source and drain voltage, while increasing the plunger gate bias to ensure single electron occupancy of the QD. The tip is moved 30~nm in the $x$-direction before the next step, in which the plunger gate bias is lowered while the tip bias is increased to transfer the electron from underneath the plunger gate to a region under the tip. Next, the source, drain, plunger and screening gates are set to equal voltages to reduce variations in the potential when guiding the electron to the ungated surface. Finally, the tip is moved 350 nm in the $x$-direction to a region far away from the patterned gate electrodes as shown in Fig.\ 2(a)(vi). During this process the tip voltage is changed to demonstrate the capability for an adiabatic evolution. To ensure adiabaticity in the simulation, we require that a single electron is present in the dot region at all positions while keeping the ground state of the electron below the Fermi level. In the simulation reported, the energy was kept constant at a point where the first excited state energy was higher than the Fermi level. Thus, we have demonstrated a physically realizable protocol to load a single electron probe dot in the QW under a movable SGM tip.

Lastly, we demonstrate how the tip-induced dot can be utilized to probe correlations between defects in the Si/SiGe heterostructure and the valley splitting. In Fig.~3, we simulate the lateral motion of the electronic wavefunction in the QW with a lattice defect situated at $y$ = 500 nm. The defect is a single atom step edge located at the interface of the Si QW and the upper Si$_{0.7}$Ge$_{0.3}$ spacer layer as shown in Fig.\ 3(a). The SGM tip is then translated in the $y$-direction in order to move the tip-induced dot across the step edge. Figure 3(a) confirms that the electron wavefunction moves with the tip as the latter is translated across the step edge. Finally, the valley splitting is calculated using a tight-binding model for different $y$-positions of the tip in Fig.\ 3(b) \cite{boykin2004valley, kharche2007valley, Jiang2012effects, Abadillo-Uriel2018signatures, EkmelPRB}. The drastic dip in valley splitting at $y_{\rm T}$ = 500 nm results from the step-edge defect. Even though recent studies have shown that the leading cause of valley splitting in Si/SiGe devices is alloy disorder \cite{Wuetz2022atomic, Losert2023practical, pena2023utilizing, volmer2023mapping, esposti2024low}, the step edge model provides an upper bound of the valley splitting \cite{Losert2023practical}. Therefore, it can be used to confirm that the change in valley splitting is well below the first orbital excitation, calculated to be 2.92 meV. This supports that SGM can be reliably utilized to detect spatial variations of the valley splitting in Si/SiGe heterostructures.

In conclusion, our simulations demonstrate the feasibility of using a Si/SiGe heterostructure with pre-fabricated gate electrodes and a scannable SGM tip to probe correlations between atomic sized defects at the Si/SiGe interface and local reductions in the valley splitting. Microwave spectroscopy has previously been utilized in gated Si/SiGe quantum devices \cite{burkard2016dispersive, mi2017high}, suggesting that readout can be performed with a superconducting microwave resonator embedded in the SGM tip. The approach investigated here may allow for more rapid characterization of the valley splitting in Si QWs, without the need to serially fabricate many single electron quantum dots.

\begin{acknowledgements}
Supported by AFOSR grant FA9550-23-1-0710, ARO grants W911NF-15-1-0149 and W911NF-23-1-0104, and the Gordon and Betty Moore Foundation Emergent Phenomena in Quantum Systems (EPiQS) Initiative through Grant No. GBMF4535.

\end{acknowledgements}

\bibliography{SGM_v4}

\begin{thebibliography}{48}%
\makeatletter
\providecommand \@ifxundefined [1]{%
 \@ifx{#1\undefined}
}%
\providecommand \@ifnum [1]{%
 \ifnum #1\expandafter \@firstoftwo
 \else \expandafter \@secondoftwo
 \fi
}%
\providecommand \@ifx [1]{%
 \ifx #1\expandafter \@firstoftwo
 \else \expandafter \@secondoftwo
 \fi
}%
\providecommand \natexlab [1]{#1}%
\providecommand \enquote  [1]{``#1''}%
\providecommand \bibnamefont  [1]{#1}%
\providecommand \bibfnamefont [1]{#1}%
\providecommand \citenamefont [1]{#1}%
\providecommand \href@noop [0]{\@secondoftwo}%
\providecommand \href [0]{\begingroup \@sanitize@url \@href}%
\providecommand \@href[1]{\@@startlink{#1}\@@href}%
\providecommand \@@href[1]{\endgroup#1\@@endlink}%
\providecommand \@sanitize@url [0]{\catcode `\\12\catcode `\$12\catcode
  `\&12\catcode `\#12\catcode `\^12\catcode `\_12\catcode `\%12\relax}%
\providecommand \@@startlink[1]{}%
\providecommand \@@endlink[0]{}%
\providecommand \url  [0]{\begingroup\@sanitize@url \@url }%
\providecommand \@url [1]{\endgroup\@href {#1}{\urlprefix }}%
\providecommand \urlprefix  [0]{URL }%
\providecommand \Eprint [0]{\href }%
\providecommand \doibase [0]{https://doi.org/}%
\providecommand \selectlanguage [0]{\@gobble}%
\providecommand \bibinfo  [0]{\@secondoftwo}%
\providecommand \bibfield  [0]{\@secondoftwo}%
\providecommand \translation [1]{[#1]}%
\providecommand \BibitemOpen [0]{}%
\providecommand \bibitemStop [0]{}%
\providecommand \bibitemNoStop [0]{.\EOS\space}%
\providecommand \EOS [0]{\spacefactor3000\relax}%
\providecommand \BibitemShut  [1]{\csname bibitem#1\endcsname}%
\let\auto@bib@innerbib\@empty
\bibitem [{\citenamefont {Burkard}\ \emph {et~al.}(2023)\citenamefont
  {Burkard}, \citenamefont {Ladd}, \citenamefont {Pan}, \citenamefont
  {Nichol},\ and\ \citenamefont {Petta}}]{burkard2023semiconductor}%
  \BibitemOpen
  \bibfield  {author} {\bibinfo {author} {\bibfnamefont {G.}~\bibnamefont
  {Burkard}}, \bibinfo {author} {\bibfnamefont {T.~D.}\ \bibnamefont {Ladd}},
  \bibinfo {author} {\bibfnamefont {A.}~\bibnamefont {Pan}}, \bibinfo {author}
  {\bibfnamefont {J.~M.}\ \bibnamefont {Nichol}},\ and\ \bibinfo {author}
  {\bibfnamefont {J.~R.}\ \bibnamefont {Petta}},\ }\bibfield  {title} {\bibinfo
  {title} {Semiconductor spin qubits},\ }\href
  {https://doi.org/10.1103/RevModPhys.95.025003} {\bibfield  {journal}
  {\bibinfo  {journal} {Rev. Mod. Phys.}\ }\textbf {\bibinfo {volume} {95}},\
  \bibinfo {pages} {025003} (\bibinfo {year} {2023})}\BibitemShut {NoStop}%
\bibitem [{\citenamefont {Tyryshkin}\ \emph {et~al.}(2012)\citenamefont
  {Tyryshkin}, \citenamefont {Tojo}, \citenamefont {Morton}, \citenamefont
  {Riemann}, \citenamefont {Abrosimov}, \citenamefont {Becker}, \citenamefont
  {Pohl}, \citenamefont {Schenkel}, \citenamefont {Thewalt}, \citenamefont
  {Itoh},\ and\ \citenamefont {Lyon}}]{tyryshkin2012electron}%
  \BibitemOpen
  \bibfield  {author} {\bibinfo {author} {\bibfnamefont {A.~M.}\ \bibnamefont
  {Tyryshkin}}, \bibinfo {author} {\bibfnamefont {S.}~\bibnamefont {Tojo}},
  \bibinfo {author} {\bibfnamefont {J.~J.~L.}\ \bibnamefont {Morton}}, \bibinfo
  {author} {\bibfnamefont {H.}~\bibnamefont {Riemann}}, \bibinfo {author}
  {\bibfnamefont {N.~V.}\ \bibnamefont {Abrosimov}}, \bibinfo {author}
  {\bibfnamefont {P.}~\bibnamefont {Becker}}, \bibinfo {author} {\bibfnamefont
  {H.-J.}\ \bibnamefont {Pohl}}, \bibinfo {author} {\bibfnamefont
  {T.}~\bibnamefont {Schenkel}}, \bibinfo {author} {\bibfnamefont {M.~L.~W.}\
  \bibnamefont {Thewalt}}, \bibinfo {author} {\bibfnamefont {K.~M.}\
  \bibnamefont {Itoh}},\ and\ \bibinfo {author} {\bibfnamefont {S.~A.}\
  \bibnamefont {Lyon}},\ }\bibfield  {title} {\bibinfo {title} {Electron spin
  coherence exceeding seconds in high-purity silicon},\ }\href
  {https://doi.org/10.1038/nmat3182} {\bibfield  {journal} {\bibinfo  {journal}
  {Nat. Mat.}\ }\textbf {\bibinfo {volume} {11}},\ \bibinfo {pages} {143}
  (\bibinfo {year} {2012})}\BibitemShut {NoStop}%
\bibitem [{\citenamefont {Mills}\ \emph {et~al.}(2022)\citenamefont {Mills},
  \citenamefont {Guinn}, \citenamefont {Gullans}, \citenamefont {Sigillito},
  \citenamefont {Feldman}, \citenamefont {Nielsen},\ and\ \citenamefont
  {Petta}}]{mills2022two}%
  \BibitemOpen
  \bibfield  {author} {\bibinfo {author} {\bibfnamefont {A.~R.}\ \bibnamefont
  {Mills}}, \bibinfo {author} {\bibfnamefont {C.~R.}\ \bibnamefont {Guinn}},
  \bibinfo {author} {\bibfnamefont {M.~J.}\ \bibnamefont {Gullans}}, \bibinfo
  {author} {\bibfnamefont {A.~J.}\ \bibnamefont {Sigillito}}, \bibinfo {author}
  {\bibfnamefont {M.~M.}\ \bibnamefont {Feldman}}, \bibinfo {author}
  {\bibfnamefont {E.}~\bibnamefont {Nielsen}},\ and\ \bibinfo {author}
  {\bibfnamefont {J.~R.}\ \bibnamefont {Petta}},\ }\bibfield  {title} {\bibinfo
  {title} {Two-qubit silicon quantum processor with operation fidelity
  exceeding 99\%},\ }\href {https://doi.org/10.1126/sciadv.abn5130} {\bibfield
  {journal} {\bibinfo  {journal} {Sci. Adv.}\ }\textbf {\bibinfo {volume}
  {8}},\ \bibinfo {pages} {eabn5130} (\bibinfo {year} {2022})}\BibitemShut
  {NoStop}%
\bibitem [{\citenamefont {Xue}\ \emph {et~al.}(2022)\citenamefont {Xue},
  \citenamefont {Russ}, \citenamefont {Samkharadze}, \citenamefont {Undseth},
  \citenamefont {Sammak}, \citenamefont {Scappucci},\ and\ \citenamefont
  {Vandersypen}}]{xue2022quantum}%
  \BibitemOpen
  \bibfield  {author} {\bibinfo {author} {\bibfnamefont {X.}~\bibnamefont
  {Xue}}, \bibinfo {author} {\bibfnamefont {M.}~\bibnamefont {Russ}}, \bibinfo
  {author} {\bibfnamefont {N.}~\bibnamefont {Samkharadze}}, \bibinfo {author}
  {\bibfnamefont {B.}~\bibnamefont {Undseth}}, \bibinfo {author} {\bibfnamefont
  {A.}~\bibnamefont {Sammak}}, \bibinfo {author} {\bibfnamefont
  {G.}~\bibnamefont {Scappucci}},\ and\ \bibinfo {author} {\bibfnamefont
  {L.~M.~K.}\ \bibnamefont {Vandersypen}},\ }\bibfield  {title} {\bibinfo
  {title} {Quantum logic with spin qubits crossing the surface code
  threshold},\ }\href {https://doi.org/10.1038/s41586-021-04273-w} {\bibfield
  {journal} {\bibinfo  {journal} {Nature}\ }\textbf {\bibinfo {volume} {601}},\
  \bibinfo {pages} {343} (\bibinfo {year} {2022})}\BibitemShut {NoStop}%
\bibitem [{\citenamefont {Noiri}\ \emph {et~al.}(2022)\citenamefont {Noiri},
  \citenamefont {Takeda}, \citenamefont {Nakajima}, \citenamefont {Kobayashi},
  \citenamefont {Sammak}, \citenamefont {Scappucci},\ and\ \citenamefont
  {Tarucha}}]{noiri2022fast}%
  \BibitemOpen
  \bibfield  {author} {\bibinfo {author} {\bibfnamefont {A.}~\bibnamefont
  {Noiri}}, \bibinfo {author} {\bibfnamefont {K.}~\bibnamefont {Takeda}},
  \bibinfo {author} {\bibfnamefont {T.}~\bibnamefont {Nakajima}}, \bibinfo
  {author} {\bibfnamefont {T.}~\bibnamefont {Kobayashi}}, \bibinfo {author}
  {\bibfnamefont {A.}~\bibnamefont {Sammak}}, \bibinfo {author} {\bibfnamefont
  {G.}~\bibnamefont {Scappucci}},\ and\ \bibinfo {author} {\bibfnamefont
  {S.}~\bibnamefont {Tarucha}},\ }\bibfield  {title} {\bibinfo {title} {Fast
  universal quantum gate above the fault-tolerance threshold in silicon},\
  }\href {https://doi.org/10.1038/s41586-021-04182-y} {\bibfield  {journal}
  {\bibinfo  {journal} {Nature}\ }\textbf {\bibinfo {volume} {601}},\ \bibinfo
  {pages} {338} (\bibinfo {year} {2022})}\BibitemShut {NoStop}%
\bibitem [{\citenamefont {Zajac}\ \emph {et~al.}(2016)\citenamefont {Zajac},
  \citenamefont {Hazard}, \citenamefont {Mi}, \citenamefont {Nielsen},\ and\
  \citenamefont {Petta}}]{zajac2016scalable}%
  \BibitemOpen
  \bibfield  {author} {\bibinfo {author} {\bibfnamefont {D.~M.}\ \bibnamefont
  {Zajac}}, \bibinfo {author} {\bibfnamefont {T.~M.}\ \bibnamefont {Hazard}},
  \bibinfo {author} {\bibfnamefont {X.}~\bibnamefont {Mi}}, \bibinfo {author}
  {\bibfnamefont {E.}~\bibnamefont {Nielsen}},\ and\ \bibinfo {author}
  {\bibfnamefont {J.~R.}\ \bibnamefont {Petta}},\ }\bibfield  {title} {\bibinfo
  {title} {Scalable gate architecture for a one-dimensional array of
  semiconductor spin qubits},\ }\href
  {https://doi.org/10.1103/PhysRevApplied.6.054013} {\bibfield  {journal}
  {\bibinfo  {journal} {Phys. Rev. Appl.}\ }\textbf {\bibinfo {volume} {6}},\
  \bibinfo {pages} {054013} (\bibinfo {year} {2016})}\BibitemShut {NoStop}%
\bibitem [{\citenamefont {Freeman}\ \emph {et~al.}(2016)\citenamefont
  {Freeman}, \citenamefont {Schoenfield},\ and\ \citenamefont
  {Jiang}}]{freeman2016comparison}%
  \BibitemOpen
  \bibfield  {author} {\bibinfo {author} {\bibfnamefont {B.~M.}\ \bibnamefont
  {Freeman}}, \bibinfo {author} {\bibfnamefont {J.~S.}\ \bibnamefont
  {Schoenfield}},\ and\ \bibinfo {author} {\bibfnamefont {H.}~\bibnamefont
  {Jiang}},\ }\bibfield  {title} {\bibinfo {title} {{Comparison of low
  frequency charge noise in identically patterned Si/SiO$_2$ and Si/SiGe
  quantum dots}},\ }\href {https://doi.org/10.1063/1.4954700} {\bibfield
  {journal} {\bibinfo  {journal} {Appl. Phys. Lett.}\ }\textbf {\bibinfo
  {volume} {108}},\ \bibinfo {pages} {253108} (\bibinfo {year}
  {2016})}\BibitemShut {NoStop}%
\bibitem [{\citenamefont {Mi}\ \emph {et~al.}(2018)\citenamefont {Mi},
  \citenamefont {Kohler},\ and\ \citenamefont {Petta}}]{mi2018landau}%
  \BibitemOpen
  \bibfield  {author} {\bibinfo {author} {\bibfnamefont {X.}~\bibnamefont
  {Mi}}, \bibinfo {author} {\bibfnamefont {S.}~\bibnamefont {Kohler}},\ and\
  \bibinfo {author} {\bibfnamefont {J.~R.}\ \bibnamefont {Petta}},\ }\bibfield
  {title} {\bibinfo {title} {{Landau-Zener interferometry of valley-orbit
  states in Si/SiGe double quantum dots}},\ }\href
  {https://doi.org/10.1103/PhysRevB.98.161404} {\bibfield  {journal} {\bibinfo
  {journal} {Phys. Rev. B}\ }\textbf {\bibinfo {volume} {98}},\ \bibinfo
  {pages} {161404} (\bibinfo {year} {2018})}\BibitemShut {NoStop}%
\bibitem [{\citenamefont {Connors}\ \emph {et~al.}(2022)\citenamefont
  {Connors}, \citenamefont {Nelson}, \citenamefont {Edge},\ and\ \citenamefont
  {Nichol}}]{connors2022charge}%
  \BibitemOpen
  \bibfield  {author} {\bibinfo {author} {\bibfnamefont {E.~J.}\ \bibnamefont
  {Connors}}, \bibinfo {author} {\bibfnamefont {J.}~\bibnamefont {Nelson}},
  \bibinfo {author} {\bibfnamefont {L.~F.}\ \bibnamefont {Edge}},\ and\
  \bibinfo {author} {\bibfnamefont {J.~M.}\ \bibnamefont {Nichol}},\ }\bibfield
   {title} {\bibinfo {title} {{Charge-noise spectroscopy of Si/SiGe quantum
  dots via dynamically-decoupled exchange oscillations}},\ }\href
  {https://doi.org/10.1038/s41467-022-28519-x} {\bibfield  {journal} {\bibinfo
  {journal} {Nat. Commun.}\ }\textbf {\bibinfo {volume} {13}},\ \bibinfo
  {pages} {940} (\bibinfo {year} {2022})}\BibitemShut {NoStop}%
\bibitem [{\citenamefont {Ando}\ \emph {et~al.}(1982)\citenamefont {Ando},
  \citenamefont {Fowler},\ and\ \citenamefont {Stern}}]{ando1982electronic}%
  \BibitemOpen
  \bibfield  {author} {\bibinfo {author} {\bibfnamefont {T.}~\bibnamefont
  {Ando}}, \bibinfo {author} {\bibfnamefont {A.~B.}\ \bibnamefont {Fowler}},\
  and\ \bibinfo {author} {\bibfnamefont {F.}~\bibnamefont {Stern}},\ }\bibfield
   {title} {\bibinfo {title} {Electronic properties of two-dimensional
  systems},\ }\href {https://doi.org/10.1103/RevModPhys.54.437} {\bibfield
  {journal} {\bibinfo  {journal} {Rev. Mod. Phys.}\ }\textbf {\bibinfo {volume}
  {54}},\ \bibinfo {pages} {437} (\bibinfo {year} {1982})}\BibitemShut
  {NoStop}%
\bibitem [{\citenamefont {Burkard}\ and\ \citenamefont
  {Petta}(2016)}]{burkard2016dispersive}%
  \BibitemOpen
  \bibfield  {author} {\bibinfo {author} {\bibfnamefont {G.}~\bibnamefont
  {Burkard}}\ and\ \bibinfo {author} {\bibfnamefont {J.~R.}\ \bibnamefont
  {Petta}},\ }\bibfield  {title} {\bibinfo {title} {Dispersive readout of
  valley splittings in cavity-coupled silicon quantum dots},\ }\href
  {https://doi.org/10.1103/PhysRevB.94.195305} {\bibfield  {journal} {\bibinfo
  {journal} {Phys. Rev. B}\ }\textbf {\bibinfo {volume} {94}},\ \bibinfo
  {pages} {195305} (\bibinfo {year} {2016})}\BibitemShut {NoStop}%
\bibitem [{\citenamefont {Schäffler}(1997)}]{schaffler1997high}%
  \BibitemOpen
  \bibfield  {author} {\bibinfo {author} {\bibfnamefont {F.}~\bibnamefont
  {Schäffler}},\ }\bibfield  {title} {\bibinfo {title} {{High-mobility Si and
  Ge structures}},\ }\href {https://doi.org/10.1088/0268-1242/12/12/001}
  {\bibfield  {journal} {\bibinfo  {journal} {Semicond. Sci. Technol.}\
  }\textbf {\bibinfo {volume} {12}},\ \bibinfo {pages} {1515} (\bibinfo {year}
  {1997})}\BibitemShut {NoStop}%
\bibitem [{\citenamefont {Zwanenburg}\ \emph {et~al.}(2013)\citenamefont
  {Zwanenburg}, \citenamefont {Dzurak}, \citenamefont {Morello}, \citenamefont
  {Simmons}, \citenamefont {Hollenberg}, \citenamefont {Klimeck}, \citenamefont
  {Rogge}, \citenamefont {Coppersmith},\ and\ \citenamefont
  {Eriksson}}]{Zwanenburg2013silicon}%
  \BibitemOpen
  \bibfield  {author} {\bibinfo {author} {\bibfnamefont {F.~A.}\ \bibnamefont
  {Zwanenburg}}, \bibinfo {author} {\bibfnamefont {A.~S.}\ \bibnamefont
  {Dzurak}}, \bibinfo {author} {\bibfnamefont {A.}~\bibnamefont {Morello}},
  \bibinfo {author} {\bibfnamefont {M.~Y.}\ \bibnamefont {Simmons}}, \bibinfo
  {author} {\bibfnamefont {L.~C.~L.}\ \bibnamefont {Hollenberg}}, \bibinfo
  {author} {\bibfnamefont {G.}~\bibnamefont {Klimeck}}, \bibinfo {author}
  {\bibfnamefont {S.}~\bibnamefont {Rogge}}, \bibinfo {author} {\bibfnamefont
  {S.~N.}\ \bibnamefont {Coppersmith}},\ and\ \bibinfo {author} {\bibfnamefont
  {M.~A.}\ \bibnamefont {Eriksson}},\ }\bibfield  {title} {\bibinfo {title}
  {Silicon quantum electronics},\ }\href
  {https://doi.org/10.1103/RevModPhys.85.961} {\bibfield  {journal} {\bibinfo
  {journal} {Rev. Mod. Phys.}\ }\textbf {\bibinfo {volume} {85}},\ \bibinfo
  {pages} {961} (\bibinfo {year} {2013})}\BibitemShut {NoStop}%
\bibitem [{\citenamefont {Zajac}\ \emph {et~al.}(2015)\citenamefont {Zajac},
  \citenamefont {Hazard}, \citenamefont {Mi}, \citenamefont {Wang},\ and\
  \citenamefont {Petta}}]{zajac2015a}%
  \BibitemOpen
  \bibfield  {author} {\bibinfo {author} {\bibfnamefont {D.~M.}\ \bibnamefont
  {Zajac}}, \bibinfo {author} {\bibfnamefont {T.~M.}\ \bibnamefont {Hazard}},
  \bibinfo {author} {\bibfnamefont {X.}~\bibnamefont {Mi}}, \bibinfo {author}
  {\bibfnamefont {K.}~\bibnamefont {Wang}},\ and\ \bibinfo {author}
  {\bibfnamefont {J.~R.}\ \bibnamefont {Petta}},\ }\bibfield  {title} {\bibinfo
  {title} {{A reconfigurable gate architecture for Si/SiGe quantum dots}},\
  }\href {https://doi.org/10.1063/1.4922249} {\bibfield  {journal} {\bibinfo
  {journal} {Appl. Phys. Lett.}\ }\textbf {\bibinfo {volume} {106}},\ \bibinfo
  {pages} {223507} (\bibinfo {year} {2015})}\BibitemShut {NoStop}%
\bibitem [{\citenamefont {Mi}\ \emph {et~al.}(2017)\citenamefont {Mi},
  \citenamefont {P\'eterfalvi}, \citenamefont {Burkard},\ and\ \citenamefont
  {Petta}}]{mi2017high}%
  \BibitemOpen
  \bibfield  {author} {\bibinfo {author} {\bibfnamefont {X.}~\bibnamefont
  {Mi}}, \bibinfo {author} {\bibfnamefont {C.~G.}\ \bibnamefont
  {P\'eterfalvi}}, \bibinfo {author} {\bibfnamefont {G.}~\bibnamefont
  {Burkard}},\ and\ \bibinfo {author} {\bibfnamefont {J.~R.}\ \bibnamefont
  {Petta}},\ }\bibfield  {title} {\bibinfo {title} {{High-Resolution Valley
  Spectroscopy of Si Quantum Dots}},\ }\href
  {https://doi.org/10.1103/PhysRevLett.119.176803} {\bibfield  {journal}
  {\bibinfo  {journal} {Phys. Rev. Lett.}\ }\textbf {\bibinfo {volume} {119}},\
  \bibinfo {pages} {176803} (\bibinfo {year} {2017})}\BibitemShut {NoStop}%
\bibitem [{\citenamefont {Hollmann}\ \emph {et~al.}(2020)\citenamefont
  {Hollmann}, \citenamefont {Struck}, \citenamefont {Langrock}, \citenamefont
  {Schmidbauer}, \citenamefont {Schauer}, \citenamefont {Leonhardt},
  \citenamefont {Sawano}, \citenamefont {Riemann}, \citenamefont {Abrosimov},
  \citenamefont {Bougeard},\ and\ \citenamefont
  {Schreiber}}]{hollman2020large}%
  \BibitemOpen
  \bibfield  {author} {\bibinfo {author} {\bibfnamefont {A.}~\bibnamefont
  {Hollmann}}, \bibinfo {author} {\bibfnamefont {T.}~\bibnamefont {Struck}},
  \bibinfo {author} {\bibfnamefont {V.}~\bibnamefont {Langrock}}, \bibinfo
  {author} {\bibfnamefont {A.}~\bibnamefont {Schmidbauer}}, \bibinfo {author}
  {\bibfnamefont {F.}~\bibnamefont {Schauer}}, \bibinfo {author} {\bibfnamefont
  {T.}~\bibnamefont {Leonhardt}}, \bibinfo {author} {\bibfnamefont
  {K.}~\bibnamefont {Sawano}}, \bibinfo {author} {\bibfnamefont
  {H.}~\bibnamefont {Riemann}}, \bibinfo {author} {\bibfnamefont {N.~V.}\
  \bibnamefont {Abrosimov}}, \bibinfo {author} {\bibfnamefont {D.}~\bibnamefont
  {Bougeard}},\ and\ \bibinfo {author} {\bibfnamefont {L.~R.}\ \bibnamefont
  {Schreiber}},\ }\bibfield  {title} {\bibinfo {title} {{Large, Tunable Valley
  Splitting and Single-Spin Relaxation Mechanisms in a Si-SiGe Quantum Dot}},\
  }\href {https://doi.org/10.1103/PhysRevApplied.13.034068} {\bibfield
  {journal} {\bibinfo  {journal} {Phys. Rev. Appl.}\ }\textbf {\bibinfo
  {volume} {13}},\ \bibinfo {pages} {034068} (\bibinfo {year}
  {2020})}\BibitemShut {NoStop}%
\bibitem [{\citenamefont {Chen}\ \emph {et~al.}(2021)\citenamefont {Chen},
  \citenamefont {Raach}, \citenamefont {Pan}, \citenamefont {Kiselev},
  \citenamefont {Acuna}, \citenamefont {Blumoff}, \citenamefont {Brecht},
  \citenamefont {Choi}, \citenamefont {Ha}, \citenamefont {Hulbert},
  \citenamefont {Jura}, \citenamefont {Keating}, \citenamefont {Noah},
  \citenamefont {Sun}, \citenamefont {Thomas}, \citenamefont {Borselli},
  \citenamefont {Jackson}, \citenamefont {Rakher},\ and\ \citenamefont
  {Ross}}]{chen2021detuning}%
  \BibitemOpen
  \bibfield  {author} {\bibinfo {author} {\bibfnamefont {E.~H.}\ \bibnamefont
  {Chen}}, \bibinfo {author} {\bibfnamefont {K.}~\bibnamefont {Raach}},
  \bibinfo {author} {\bibfnamefont {A.}~\bibnamefont {Pan}}, \bibinfo {author}
  {\bibfnamefont {A.~A.}\ \bibnamefont {Kiselev}}, \bibinfo {author}
  {\bibfnamefont {E.}~\bibnamefont {Acuna}}, \bibinfo {author} {\bibfnamefont
  {J.~Z.}\ \bibnamefont {Blumoff}}, \bibinfo {author} {\bibfnamefont
  {T.}~\bibnamefont {Brecht}}, \bibinfo {author} {\bibfnamefont {M.~D.}\
  \bibnamefont {Choi}}, \bibinfo {author} {\bibfnamefont {W.}~\bibnamefont
  {Ha}}, \bibinfo {author} {\bibfnamefont {D.~R.}\ \bibnamefont {Hulbert}},
  \bibinfo {author} {\bibfnamefont {M.~P.}\ \bibnamefont {Jura}}, \bibinfo
  {author} {\bibfnamefont {T.~E.}\ \bibnamefont {Keating}}, \bibinfo {author}
  {\bibfnamefont {R.}~\bibnamefont {Noah}}, \bibinfo {author} {\bibfnamefont
  {B.}~\bibnamefont {Sun}}, \bibinfo {author} {\bibfnamefont {B.~J.}\
  \bibnamefont {Thomas}}, \bibinfo {author} {\bibfnamefont {M.~G.}\
  \bibnamefont {Borselli}}, \bibinfo {author} {\bibfnamefont {C.}~\bibnamefont
  {Jackson}}, \bibinfo {author} {\bibfnamefont {M.~T.}\ \bibnamefont
  {Rakher}},\ and\ \bibinfo {author} {\bibfnamefont {R.~S.}\ \bibnamefont
  {Ross}},\ }\bibfield  {title} {\bibinfo {title} {Detuning axis pulsed
  spectroscopy of valley-orbital states in
  $\mathrm{Si}$/$\mathrm{Si}$-$\mathrm{Ge}$ quantum dots},\ }\href
  {https://doi.org/10.1103/PhysRevApplied.15.044033} {\bibfield  {journal}
  {\bibinfo  {journal} {Phys. Rev. Appl.}\ }\textbf {\bibinfo {volume} {15}},\
  \bibinfo {pages} {044033} (\bibinfo {year} {2021})}\BibitemShut {NoStop}%
\bibitem [{\citenamefont {Borselli}\ \emph {et~al.}(2011)\citenamefont
  {Borselli}, \citenamefont {Ross}, \citenamefont {Kiselev}, \citenamefont
  {Croke}, \citenamefont {Holabird}, \citenamefont {Deelman}, \citenamefont
  {Warren}, \citenamefont {Alvarado-Rodriguez}, \citenamefont {Milosavljevic},
  \citenamefont {Ku}, \citenamefont {Wong}, \citenamefont {Schmitz},
  \citenamefont {Sokolich}, \citenamefont {Gyure},\ and\ \citenamefont
  {Hunter}}]{Borselli2011measurement}%
  \BibitemOpen
  \bibfield  {author} {\bibinfo {author} {\bibfnamefont {M.~G.}\ \bibnamefont
  {Borselli}}, \bibinfo {author} {\bibfnamefont {R.~S.}\ \bibnamefont {Ross}},
  \bibinfo {author} {\bibfnamefont {A.~A.}\ \bibnamefont {Kiselev}}, \bibinfo
  {author} {\bibfnamefont {E.~T.}\ \bibnamefont {Croke}}, \bibinfo {author}
  {\bibfnamefont {K.~S.}\ \bibnamefont {Holabird}}, \bibinfo {author}
  {\bibfnamefont {P.~W.}\ \bibnamefont {Deelman}}, \bibinfo {author}
  {\bibfnamefont {L.~D.}\ \bibnamefont {Warren}}, \bibinfo {author}
  {\bibfnamefont {I.}~\bibnamefont {Alvarado-Rodriguez}}, \bibinfo {author}
  {\bibfnamefont {I.}~\bibnamefont {Milosavljevic}}, \bibinfo {author}
  {\bibfnamefont {F.~C.}\ \bibnamefont {Ku}}, \bibinfo {author} {\bibfnamefont
  {W.~S.}\ \bibnamefont {Wong}}, \bibinfo {author} {\bibfnamefont {A.~E.}\
  \bibnamefont {Schmitz}}, \bibinfo {author} {\bibfnamefont {M.}~\bibnamefont
  {Sokolich}}, \bibinfo {author} {\bibfnamefont {M.~F.}\ \bibnamefont
  {Gyure}},\ and\ \bibinfo {author} {\bibfnamefont {A.~T.}\ \bibnamefont
  {Hunter}},\ }\bibfield  {title} {\bibinfo {title} {{Measurement of valley
  splitting in high-symmetry Si/SiGe quantum dots}},\ }\href
  {https://doi.org/10.1063/1.3569717} {\bibfield  {journal} {\bibinfo
  {journal} {Appl. Phys. Lett.}\ }\textbf {\bibinfo {volume} {98}},\ \bibinfo
  {pages} {123118} (\bibinfo {year} {2011})}\BibitemShut {NoStop}%
\bibitem [{\citenamefont {Schoenfield}\ \emph {et~al.}(2017)\citenamefont
  {Schoenfield}, \citenamefont {Freeman},\ and\ \citenamefont
  {Jiang}}]{Schoenfield2017coherent}%
  \BibitemOpen
  \bibfield  {author} {\bibinfo {author} {\bibfnamefont {J.~S.}\ \bibnamefont
  {Schoenfield}}, \bibinfo {author} {\bibfnamefont {B.~M.}\ \bibnamefont
  {Freeman}},\ and\ \bibinfo {author} {\bibfnamefont {H.}~\bibnamefont
  {Jiang}},\ }\bibfield  {title} {\bibinfo {title} {Coherent manipulation of
  valley states at multiple charge configurations of a silicon quantum dot
  device},\ }\href {https://doi.org/10.1038/s41467-017-00073-x} {\bibfield
  {journal} {\bibinfo  {journal} {Nat. Commun.}\ }\textbf {\bibinfo {volume}
  {8}},\ \bibinfo {pages} {64} (\bibinfo {year} {2017})}\BibitemShut {NoStop}%
\bibitem [{\citenamefont {Friesen}\ \emph {et~al.}(2006)\citenamefont
  {Friesen}, \citenamefont {Eriksson},\ and\ \citenamefont
  {Coppersmith}}]{friesen2006magnetic}%
  \BibitemOpen
  \bibfield  {author} {\bibinfo {author} {\bibfnamefont {M.}~\bibnamefont
  {Friesen}}, \bibinfo {author} {\bibfnamefont {M.~A.}\ \bibnamefont
  {Eriksson}},\ and\ \bibinfo {author} {\bibfnamefont {S.~N.}\ \bibnamefont
  {Coppersmith}},\ }\bibfield  {title} {\bibinfo {title} {{Magnetic field
  dependence of valley splitting in realistic Si-SiGe quantum wells}},\ }\href
  {https://doi.org/10.1063/1.2387975} {\bibfield  {journal} {\bibinfo
  {journal} {Appl. Phys. Lett.}\ }\textbf {\bibinfo {volume} {89}},\ \bibinfo
  {pages} {202106} (\bibinfo {year} {2006})}\BibitemShut {NoStop}%
\bibitem [{\citenamefont {Kharche}\ \emph {et~al.}(2007)\citenamefont
  {Kharche}, \citenamefont {Prada}, \citenamefont {Boykin},\ and\ \citenamefont
  {Klimeck}}]{kharche2007valley}%
  \BibitemOpen
  \bibfield  {author} {\bibinfo {author} {\bibfnamefont {N.}~\bibnamefont
  {Kharche}}, \bibinfo {author} {\bibfnamefont {M.}~\bibnamefont {Prada}},
  \bibinfo {author} {\bibfnamefont {T.~B.}\ \bibnamefont {Boykin}},\ and\
  \bibinfo {author} {\bibfnamefont {G.}~\bibnamefont {Klimeck}},\ }\bibfield
  {title} {\bibinfo {title} {{Valley splitting in strained silicon quantum
  wells modeled with 2° miscuts, step disorder, and alloy disorder}},\ }\href
  {https://doi.org/10.1063/1.2591432} {\bibfield  {journal} {\bibinfo
  {journal} {Appl. Phys. Lett.}\ }\textbf {\bibinfo {volume} {90}},\ \bibinfo
  {pages} {092109} (\bibinfo {year} {2007})}\BibitemShut {NoStop}%
\bibitem [{\citenamefont {Culcer}\ \emph {et~al.}(2010)\citenamefont {Culcer},
  \citenamefont {Hu},\ and\ \citenamefont {Das~Sarma}}]{culcer2010interface}%
  \BibitemOpen
  \bibfield  {author} {\bibinfo {author} {\bibfnamefont {D.}~\bibnamefont
  {Culcer}}, \bibinfo {author} {\bibfnamefont {X.}~\bibnamefont {Hu}},\ and\
  \bibinfo {author} {\bibfnamefont {S.}~\bibnamefont {Das~Sarma}},\ }\bibfield
  {title} {\bibinfo {title} {Interface roughness, valley-orbit coupling, and
  valley manipulation in quantum dots},\ }\href
  {https://doi.org/10.1103/PhysRevB.82.205315} {\bibfield  {journal} {\bibinfo
  {journal} {Phys. Rev. B}\ }\textbf {\bibinfo {volume} {82}},\ \bibinfo
  {pages} {205315} (\bibinfo {year} {2010})}\BibitemShut {NoStop}%
\bibitem [{\citenamefont {Gamble}\ \emph {et~al.}(2013)\citenamefont {Gamble},
  \citenamefont {Eriksson}, \citenamefont {Coppersmith},\ and\ \citenamefont
  {Friesen}}]{gamble2013disorder}%
  \BibitemOpen
  \bibfield  {author} {\bibinfo {author} {\bibfnamefont {J.~K.}\ \bibnamefont
  {Gamble}}, \bibinfo {author} {\bibfnamefont {M.~A.}\ \bibnamefont
  {Eriksson}}, \bibinfo {author} {\bibfnamefont {S.~N.}\ \bibnamefont
  {Coppersmith}},\ and\ \bibinfo {author} {\bibfnamefont {M.}~\bibnamefont
  {Friesen}},\ }\bibfield  {title} {\bibinfo {title} {{Disorder-induced
  valley-orbit hybrid states in Si quantum dots}},\ }\href
  {https://doi.org/10.1103/PhysRevB.88.035310} {\bibfield  {journal} {\bibinfo
  {journal} {Phys. Rev. B}\ }\textbf {\bibinfo {volume} {88}},\ \bibinfo
  {pages} {035310} (\bibinfo {year} {2013})}\BibitemShut {NoStop}%
\bibitem [{\citenamefont {Boross}\ \emph {et~al.}(2016)\citenamefont {Boross},
  \citenamefont {Sz\'echenyi}, \citenamefont {Culcer},\ and\ \citenamefont
  {P\'alyi}}]{boross2016control}%
  \BibitemOpen
  \bibfield  {author} {\bibinfo {author} {\bibfnamefont {P.}~\bibnamefont
  {Boross}}, \bibinfo {author} {\bibfnamefont {G.}~\bibnamefont {Sz\'echenyi}},
  \bibinfo {author} {\bibfnamefont {D.}~\bibnamefont {Culcer}},\ and\ \bibinfo
  {author} {\bibfnamefont {A.}~\bibnamefont {P\'alyi}},\ }\bibfield  {title}
  {\bibinfo {title} {Control of valley dynamics in silicon quantum dots in the
  presence of an interface step},\ }\href
  {https://doi.org/10.1103/PhysRevB.94.035438} {\bibfield  {journal} {\bibinfo
  {journal} {Phys. Rev. B}\ }\textbf {\bibinfo {volume} {94}},\ \bibinfo
  {pages} {035438} (\bibinfo {year} {2016})}\BibitemShut {NoStop}%
\bibitem [{\citenamefont {Hosseinkhani}\ and\ \citenamefont
  {Burkard}(2020)}]{hosseinkhani2020electromagnetic}%
  \BibitemOpen
  \bibfield  {author} {\bibinfo {author} {\bibfnamefont {A.}~\bibnamefont
  {Hosseinkhani}}\ and\ \bibinfo {author} {\bibfnamefont {G.}~\bibnamefont
  {Burkard}},\ }\bibfield  {title} {\bibinfo {title} {{Electromagnetic control
  of valley splitting in ideal and disordered Si quantum dots}},\ }\href
  {https://doi.org/10.1103/PhysRevResearch.2.043180} {\bibfield  {journal}
  {\bibinfo  {journal} {Phys. Rev. Res.}\ }\textbf {\bibinfo {volume} {2}},\
  \bibinfo {pages} {043180} (\bibinfo {year} {2020})}\BibitemShut {NoStop}%
\bibitem [{\citenamefont {Dodson}\ \emph {et~al.}(2022)\citenamefont {Dodson},
  \citenamefont {Ercan}, \citenamefont {Corrigan}, \citenamefont {Losert},
  \citenamefont {Holman}, \citenamefont {McJunkin}, \citenamefont {Edge},
  \citenamefont {Friesen}, \citenamefont {Coppersmith},\ and\ \citenamefont
  {Eriksson}}]{dodson2022how}%
  \BibitemOpen
  \bibfield  {author} {\bibinfo {author} {\bibfnamefont {J.~P.}\ \bibnamefont
  {Dodson}}, \bibinfo {author} {\bibfnamefont {H.~E.}\ \bibnamefont {Ercan}},
  \bibinfo {author} {\bibfnamefont {J.}~\bibnamefont {Corrigan}}, \bibinfo
  {author} {\bibfnamefont {M.~P.}\ \bibnamefont {Losert}}, \bibinfo {author}
  {\bibfnamefont {N.}~\bibnamefont {Holman}}, \bibinfo {author} {\bibfnamefont
  {T.}~\bibnamefont {McJunkin}}, \bibinfo {author} {\bibfnamefont {L.~F.}\
  \bibnamefont {Edge}}, \bibinfo {author} {\bibfnamefont {M.}~\bibnamefont
  {Friesen}}, \bibinfo {author} {\bibfnamefont {S.~N.}\ \bibnamefont
  {Coppersmith}},\ and\ \bibinfo {author} {\bibfnamefont {M.~A.}\ \bibnamefont
  {Eriksson}},\ }\bibfield  {title} {\bibinfo {title} {How valley-orbit states
  in silicon quantum dots probe quantum well interfaces},\ }\href
  {https://doi.org/10.1103/PhysRevLett.128.146802} {\bibfield  {journal}
  {\bibinfo  {journal} {Phys. Rev. Lett.}\ }\textbf {\bibinfo {volume} {128}},\
  \bibinfo {pages} {146802} (\bibinfo {year} {2022})}\BibitemShut {NoStop}%
\bibitem [{\citenamefont {Ercan}\ \emph {et~al.}(2021)\citenamefont {Ercan},
  \citenamefont {Coppersmith},\ and\ \citenamefont {Friesen}}]{EkmelPRB}%
  \BibitemOpen
  \bibfield  {author} {\bibinfo {author} {\bibfnamefont {H.~E.}\ \bibnamefont
  {Ercan}}, \bibinfo {author} {\bibfnamefont {S.~N.}\ \bibnamefont
  {Coppersmith}},\ and\ \bibinfo {author} {\bibfnamefont {M.}~\bibnamefont
  {Friesen}},\ }\bibfield  {title} {\bibinfo {title} {{Strong electron-electron
  interactions in Si/SiGe quantum dots}},\ }\href
  {https://doi.org/10.1103/PhysRevB.104.235302} {\bibfield  {journal} {\bibinfo
   {journal} {Phys. Rev. B}\ }\textbf {\bibinfo {volume} {104}},\ \bibinfo
  {pages} {235302} (\bibinfo {year} {2021})}\BibitemShut {NoStop}%
\bibitem [{\citenamefont {Paquelet~Wuetz}\ \emph {et~al.}(2022)\citenamefont
  {Paquelet~Wuetz}, \citenamefont {Losert}, \citenamefont {Koelling},
  \citenamefont {Stehouwer}, \citenamefont {Zwerver}, \citenamefont {Philips},
  \citenamefont {Mądzik}, \citenamefont {Xue}, \citenamefont {Zheng},
  \citenamefont {Lodari}, \citenamefont {Amitonov}, \citenamefont
  {Samkharadze}, \citenamefont {Sammak}, \citenamefont {Vandersypen},
  \citenamefont {Rahman}, \citenamefont {Coppersmith}, \citenamefont
  {Moutanabbir}, \citenamefont {Friesen},\ and\ \citenamefont
  {Scappucci}}]{Wuetz2022atomic}%
  \BibitemOpen
  \bibfield  {author} {\bibinfo {author} {\bibfnamefont {B.}~\bibnamefont
  {Paquelet~Wuetz}}, \bibinfo {author} {\bibfnamefont {M.~P.}\ \bibnamefont
  {Losert}}, \bibinfo {author} {\bibfnamefont {S.}~\bibnamefont {Koelling}},
  \bibinfo {author} {\bibfnamefont {L.~E.~A.}\ \bibnamefont {Stehouwer}},
  \bibinfo {author} {\bibfnamefont {A.-M.~J.}\ \bibnamefont {Zwerver}},
  \bibinfo {author} {\bibfnamefont {S.~G.~J.}\ \bibnamefont {Philips}},
  \bibinfo {author} {\bibfnamefont {M.~T.}\ \bibnamefont {Mądzik}}, \bibinfo
  {author} {\bibfnamefont {X.}~\bibnamefont {Xue}}, \bibinfo {author}
  {\bibfnamefont {G.}~\bibnamefont {Zheng}}, \bibinfo {author} {\bibfnamefont
  {M.}~\bibnamefont {Lodari}}, \bibinfo {author} {\bibfnamefont {S.~V.}\
  \bibnamefont {Amitonov}}, \bibinfo {author} {\bibfnamefont {N.}~\bibnamefont
  {Samkharadze}}, \bibinfo {author} {\bibfnamefont {A.}~\bibnamefont {Sammak}},
  \bibinfo {author} {\bibfnamefont {L.~M.~K.}\ \bibnamefont {Vandersypen}},
  \bibinfo {author} {\bibfnamefont {R.}~\bibnamefont {Rahman}}, \bibinfo
  {author} {\bibfnamefont {S.~N.}\ \bibnamefont {Coppersmith}}, \bibinfo
  {author} {\bibfnamefont {O.}~\bibnamefont {Moutanabbir}}, \bibinfo {author}
  {\bibfnamefont {M.}~\bibnamefont {Friesen}},\ and\ \bibinfo {author}
  {\bibfnamefont {G.}~\bibnamefont {Scappucci}},\ }\bibfield  {title} {\bibinfo
  {title} {{Atomic fluctuations lifting the energy degeneracy in Si/SiGe
  quantum dots}},\ }\href {https://doi.org/10.1038/s41467-022-35458-0}
  {\bibfield  {journal} {\bibinfo  {journal} {Nat. Commun.}\ }\textbf {\bibinfo
  {volume} {13}},\ \bibinfo {pages} {7730} (\bibinfo {year}
  {2022})}\BibitemShut {NoStop}%
\bibitem [{\citenamefont {Losert}\ \emph {et~al.}(2023)\citenamefont {Losert},
  \citenamefont {Eriksson}, \citenamefont {Joynt}, \citenamefont {Rahman},
  \citenamefont {Scappucci}, \citenamefont {Coppersmith},\ and\ \citenamefont
  {Friesen}}]{Losert2023practical}%
  \BibitemOpen
  \bibfield  {author} {\bibinfo {author} {\bibfnamefont {M.~P.}\ \bibnamefont
  {Losert}}, \bibinfo {author} {\bibfnamefont {M.~A.}\ \bibnamefont
  {Eriksson}}, \bibinfo {author} {\bibfnamefont {R.}~\bibnamefont {Joynt}},
  \bibinfo {author} {\bibfnamefont {R.}~\bibnamefont {Rahman}}, \bibinfo
  {author} {\bibfnamefont {G.}~\bibnamefont {Scappucci}}, \bibinfo {author}
  {\bibfnamefont {S.~N.}\ \bibnamefont {Coppersmith}},\ and\ \bibinfo {author}
  {\bibfnamefont {M.}~\bibnamefont {Friesen}},\ }\bibfield  {title} {\bibinfo
  {title} {{Practical strategies for enhancing the valley splitting in Si/SiGe
  quantum wells}},\ }\href {https://doi.org/10.1103/PhysRevB.108.125405}
  {\bibfield  {journal} {\bibinfo  {journal} {Phys. Rev. B}\ }\textbf {\bibinfo
  {volume} {108}},\ \bibinfo {pages} {125405} (\bibinfo {year}
  {2023})}\BibitemShut {NoStop}%
\bibitem [{\citenamefont {Peña}\ \emph {et~al.}(2023)\citenamefont {Peña},
  \citenamefont {Koepke}, \citenamefont {Dycus}, \citenamefont {Mounce},
  \citenamefont {Baczewski}, \citenamefont {Jacobson},\ and\ \citenamefont
  {Bussmann}}]{pena2023utilizing}%
  \BibitemOpen
  \bibfield  {author} {\bibinfo {author} {\bibfnamefont {L.~F.}\ \bibnamefont
  {Peña}}, \bibinfo {author} {\bibfnamefont {J.~C.}\ \bibnamefont {Koepke}},
  \bibinfo {author} {\bibfnamefont {J.~H.}\ \bibnamefont {Dycus}}, \bibinfo
  {author} {\bibfnamefont {A.}~\bibnamefont {Mounce}}, \bibinfo {author}
  {\bibfnamefont {A.~D.}\ \bibnamefont {Baczewski}}, \bibinfo {author}
  {\bibfnamefont {N.~T.}\ \bibnamefont {Jacobson}},\ and\ \bibinfo {author}
  {\bibfnamefont {E.}~\bibnamefont {Bussmann}},\ }\bibfield  {title} {\bibinfo
  {title} {{Utilizing multimodal microscopy to reconstruct Si/SiGe interfacial
  atomic disorder and infer its impacts on qubit variability}},\ }\href@noop {}
  {\  (\bibinfo {year} {2023})},\ \Eprint {https://arxiv.org/abs/2306.15646}
  {arXiv:2306.15646} \BibitemShut {NoStop}%
\bibitem [{\citenamefont {Volmer}\ \emph {et~al.}(2023)\citenamefont {Volmer},
  \citenamefont {Struck}, \citenamefont {Sala}, \citenamefont {Chen},
  \citenamefont {Oberländer}, \citenamefont {Offermann}, \citenamefont {Xue},
  \citenamefont {Visser}, \citenamefont {Tu}, \citenamefont {Trellenkamp},
  \citenamefont {Łukasz Cywiński}, \citenamefont {Bluhm},\ and\ \citenamefont
  {Schreiber}}]{volmer2023mapping}%
  \BibitemOpen
  \bibfield  {author} {\bibinfo {author} {\bibfnamefont {M.}~\bibnamefont
  {Volmer}}, \bibinfo {author} {\bibfnamefont {T.}~\bibnamefont {Struck}},
  \bibinfo {author} {\bibfnamefont {A.}~\bibnamefont {Sala}}, \bibinfo {author}
  {\bibfnamefont {B.}~\bibnamefont {Chen}}, \bibinfo {author} {\bibfnamefont
  {M.}~\bibnamefont {Oberländer}}, \bibinfo {author} {\bibfnamefont
  {T.}~\bibnamefont {Offermann}}, \bibinfo {author} {\bibfnamefont
  {R.}~\bibnamefont {Xue}}, \bibinfo {author} {\bibfnamefont {L.}~\bibnamefont
  {Visser}}, \bibinfo {author} {\bibfnamefont {J.-S.}\ \bibnamefont {Tu}},
  \bibinfo {author} {\bibfnamefont {S.}~\bibnamefont {Trellenkamp}}, \bibinfo
  {author} {\bibnamefont {Łukasz Cywiński}}, \bibinfo {author} {\bibfnamefont
  {H.}~\bibnamefont {Bluhm}},\ and\ \bibinfo {author} {\bibfnamefont {L.~R.}\
  \bibnamefont {Schreiber}},\ }\bibfield  {title} {\bibinfo {title} {Mapping of
  valley-splitting by conveyor-mode spin-coherent electron shuttling},\
  }\href@noop {} {\  (\bibinfo {year} {2023})},\ \Eprint
  {https://arxiv.org/abs/2312.17694} {arXiv:2312.17694} \BibitemShut {NoStop}%
\bibitem [{\citenamefont {Esposti}\ \emph {et~al.}(2024)\citenamefont
  {Esposti}, \citenamefont {Stehouwer}, \citenamefont {Önder Gül},
  \citenamefont {Samkharadze}, \citenamefont {Déprez}, \citenamefont {Meyer},
  \citenamefont {Meijer}, \citenamefont {Tryputen}, \citenamefont {Karwal},
  \citenamefont {Botifoll}, \citenamefont {Arbiol}, \citenamefont {Amitonov},
  \citenamefont {Vandersypen}, \citenamefont {Sammak}, \citenamefont
  {Veldhorst},\ and\ \citenamefont {Scappucci}}]{esposti2024low}%
  \BibitemOpen
  \bibfield  {author} {\bibinfo {author} {\bibfnamefont {D.~D.}\ \bibnamefont
  {Esposti}}, \bibinfo {author} {\bibfnamefont {L.~E.~A.}\ \bibnamefont
  {Stehouwer}}, \bibinfo {author} {\bibnamefont {Önder Gül}}, \bibinfo
  {author} {\bibfnamefont {N.}~\bibnamefont {Samkharadze}}, \bibinfo {author}
  {\bibfnamefont {C.}~\bibnamefont {Déprez}}, \bibinfo {author} {\bibfnamefont
  {M.}~\bibnamefont {Meyer}}, \bibinfo {author} {\bibfnamefont {I.~N.}\
  \bibnamefont {Meijer}}, \bibinfo {author} {\bibfnamefont {L.}~\bibnamefont
  {Tryputen}}, \bibinfo {author} {\bibfnamefont {S.}~\bibnamefont {Karwal}},
  \bibinfo {author} {\bibfnamefont {M.}~\bibnamefont {Botifoll}}, \bibinfo
  {author} {\bibfnamefont {J.}~\bibnamefont {Arbiol}}, \bibinfo {author}
  {\bibfnamefont {S.~V.}\ \bibnamefont {Amitonov}}, \bibinfo {author}
  {\bibfnamefont {L.~M.~K.}\ \bibnamefont {Vandersypen}}, \bibinfo {author}
  {\bibfnamefont {A.}~\bibnamefont {Sammak}}, \bibinfo {author} {\bibfnamefont
  {M.}~\bibnamefont {Veldhorst}},\ and\ \bibinfo {author} {\bibfnamefont
  {G.}~\bibnamefont {Scappucci}},\ }\bibfield  {title} {\bibinfo {title} {Low
  disorder and high valley splitting in silicon},\ }\href@noop {} {\  (\bibinfo
  {year} {2024})},\ \Eprint {https://arxiv.org/abs/2309.02832}
  {arXiv:2309.02832} \BibitemShut {NoStop}%
\bibitem [{\citenamefont {Kanaar}\ \emph {et~al.}(2024)\citenamefont {Kanaar},
  \citenamefont {Ercan}, \citenamefont {Gyure},\ and\ \citenamefont
  {Kestner}}]{EkmelarXiv}%
  \BibitemOpen
  \bibfield  {author} {\bibinfo {author} {\bibfnamefont {D.~W.}\ \bibnamefont
  {Kanaar}}, \bibinfo {author} {\bibfnamefont {H.~E.}\ \bibnamefont {Ercan}},
  \bibinfo {author} {\bibfnamefont {M.~F.}\ \bibnamefont {Gyure}},\ and\
  \bibinfo {author} {\bibfnamefont {J.~P.}\ \bibnamefont {Kestner}},\
  }\bibfield  {title} {\bibinfo {title} {{Proposed real-time charge noise
  measurement via valley state reflectometry}},\ }\href@noop {} {\  (\bibinfo
  {year} {2024})},\ \Eprint {https://arxiv.org/abs/2402.14765}
  {arXiv:2402.14765} \BibitemShut {NoStop}%
\bibitem [{\citenamefont {Shim}\ \emph {et~al.}(2019)\citenamefont {Shim},
  \citenamefont {Ruskov}, \citenamefont {Hurst},\ and\ \citenamefont
  {Tahan}}]{shim2019induced}%
  \BibitemOpen
  \bibfield  {author} {\bibinfo {author} {\bibfnamefont {Y.-P.}\ \bibnamefont
  {Shim}}, \bibinfo {author} {\bibfnamefont {R.}~\bibnamefont {Ruskov}},
  \bibinfo {author} {\bibfnamefont {H.~M.}\ \bibnamefont {Hurst}},\ and\
  \bibinfo {author} {\bibfnamefont {C.}~\bibnamefont {Tahan}},\ }\bibfield
  {title} {\bibinfo {title} {{Induced quantum dot probe for material
  characterization}},\ }\href {https://doi.org/10.1063/1.5053756} {\bibfield
  {journal} {\bibinfo  {journal} {Appl. Phys. Lett.}\ }\textbf {\bibinfo
  {volume} {114}},\ \bibinfo {pages} {152105} (\bibinfo {year}
  {2019})}\BibitemShut {NoStop}%
\bibitem [{\citenamefont {Voisin}\ \emph {et~al.}(2020)\citenamefont {Voisin},
  \citenamefont {Bocquel}, \citenamefont {Tankasala}, \citenamefont {Usman},
  \citenamefont {Salfi}, \citenamefont {Rahman}, \citenamefont {Simmons},
  \citenamefont {Hollenberg},\ and\ \citenamefont {Rogge}}]{voison2020valley}%
  \BibitemOpen
  \bibfield  {author} {\bibinfo {author} {\bibfnamefont {B.}~\bibnamefont
  {Voisin}}, \bibinfo {author} {\bibfnamefont {J.}~\bibnamefont {Bocquel}},
  \bibinfo {author} {\bibfnamefont {A.}~\bibnamefont {Tankasala}}, \bibinfo
  {author} {\bibfnamefont {M.}~\bibnamefont {Usman}}, \bibinfo {author}
  {\bibfnamefont {J.}~\bibnamefont {Salfi}}, \bibinfo {author} {\bibfnamefont
  {R.}~\bibnamefont {Rahman}}, \bibinfo {author} {\bibfnamefont {M.~Y.}\
  \bibnamefont {Simmons}}, \bibinfo {author} {\bibfnamefont {L.~C.~L.}\
  \bibnamefont {Hollenberg}},\ and\ \bibinfo {author} {\bibfnamefont
  {S.}~\bibnamefont {Rogge}},\ }\bibfield  {title} {\bibinfo {title} {Valley
  interference and spin exchange at the atomic scale in silicon},\ }\href
  {https://doi.org/10.1038/s41467-020-19835-1} {\bibfield  {journal} {\bibinfo
  {journal} {Nat. Commun.}\ }\textbf {\bibinfo {volume} {11}},\ \bibinfo
  {pages} {6124} (\bibinfo {year} {2020})}\BibitemShut {NoStop}%
\bibitem [{\citenamefont {Denisov}\ \emph {et~al.}(2022)\citenamefont
  {Denisov}, \citenamefont {Oh}, \citenamefont {Fuchs}, \citenamefont {Mills},
  \citenamefont {Chen}, \citenamefont {Anderson}, \citenamefont {Gyure},
  \citenamefont {Barnard},\ and\ \citenamefont {Petta}}]{Denisov2022microwave}%
  \BibitemOpen
  \bibfield  {author} {\bibinfo {author} {\bibfnamefont {A.~O.}\ \bibnamefont
  {Denisov}}, \bibinfo {author} {\bibfnamefont {S.~W.}\ \bibnamefont {Oh}},
  \bibinfo {author} {\bibfnamefont {G.}~\bibnamefont {Fuchs}}, \bibinfo
  {author} {\bibfnamefont {A.~R.}\ \bibnamefont {Mills}}, \bibinfo {author}
  {\bibfnamefont {P.}~\bibnamefont {Chen}}, \bibinfo {author} {\bibfnamefont
  {C.~R.}\ \bibnamefont {Anderson}}, \bibinfo {author} {\bibfnamefont {M.~F.}\
  \bibnamefont {Gyure}}, \bibinfo {author} {\bibfnamefont {A.~W.}\ \bibnamefont
  {Barnard}},\ and\ \bibinfo {author} {\bibfnamefont {J.~R.}\ \bibnamefont
  {Petta}},\ }\bibfield  {title} {\bibinfo {title} {{Microwave-Frequency
  Scanning Gate Microscopy of a Si/SiGe Double Quantum Dot}},\ }\href
  {https://doi.org/10.1021/acs.nanolett.2c01098} {\bibfield  {journal}
  {\bibinfo  {journal} {Nano Lett.}\ }\textbf {\bibinfo {volume} {22}},\
  \bibinfo {pages} {4807} (\bibinfo {year} {2022})}\BibitemShut {NoStop}%
\bibitem [{\citenamefont {Denisov}\ \emph {et~al.}(2023)\citenamefont
  {Denisov}, \citenamefont {Fuchs}, \citenamefont {Oh},\ and\ \citenamefont
  {Petta}}]{denisov2023dispersive}%
  \BibitemOpen
  \bibfield  {author} {\bibinfo {author} {\bibfnamefont {A.~O.}\ \bibnamefont
  {Denisov}}, \bibinfo {author} {\bibfnamefont {G.}~\bibnamefont {Fuchs}},
  \bibinfo {author} {\bibfnamefont {S.~W.}\ \bibnamefont {Oh}},\ and\ \bibinfo
  {author} {\bibfnamefont {J.~R.}\ \bibnamefont {Petta}},\ }\bibfield  {title}
  {\bibinfo {title} {{Dispersive readout of a silicon quantum device using an
  atomic force microscope-based rf gate sensor}},\ }\href
  {https://doi.org/10.1063/5.0158196} {\bibfield  {journal} {\bibinfo
  {journal} {Appl. Phys. Lett.}\ }\textbf {\bibinfo {volume} {123}},\ \bibinfo
  {pages} {093502} (\bibinfo {year} {2023})}\BibitemShut {NoStop}%
\bibitem [{\citenamefont {Chesshire}\ and\ \citenamefont
  {Henshaw}(1990)}]{chesshire1990composite}%
  \BibitemOpen
  \bibfield  {author} {\bibinfo {author} {\bibfnamefont {G.}~\bibnamefont
  {Chesshire}}\ and\ \bibinfo {author} {\bibfnamefont {W.}~\bibnamefont
  {Henshaw}},\ }\bibfield  {title} {\bibinfo {title} {Composite overlapping
  meshes for the solution of partial differential equations},\ }\href
  {https://doi.org/https://doi.org/10.1016/0021-9991(90)90196-8} {\bibfield
  {journal} {\bibinfo  {journal} {J. Comput. Phys.}\ }\textbf {\bibinfo
  {volume} {90}},\ \bibinfo {pages} {1} (\bibinfo {year} {1990})}\BibitemShut
  {NoStop}%
\bibitem [{\citenamefont {Anderson}\ \emph {et~al.}(2022)\citenamefont
  {Anderson}, \citenamefont {Gyure}, \citenamefont {Quinn}, \citenamefont
  {Pan}, \citenamefont {Ross},\ and\ \citenamefont
  {Kiselev}}]{anderson2022high}%
  \BibitemOpen
  \bibfield  {author} {\bibinfo {author} {\bibfnamefont {C.~R.}\ \bibnamefont
  {Anderson}}, \bibinfo {author} {\bibfnamefont {M.~F.}\ \bibnamefont {Gyure}},
  \bibinfo {author} {\bibfnamefont {S.}~\bibnamefont {Quinn}}, \bibinfo
  {author} {\bibfnamefont {A.}~\bibnamefont {Pan}}, \bibinfo {author}
  {\bibfnamefont {R.~S.}\ \bibnamefont {Ross}},\ and\ \bibinfo {author}
  {\bibfnamefont {A.~A.}\ \bibnamefont {Kiselev}},\ }\bibfield  {title}
  {\bibinfo {title} {{High-precision real-space simulation of electrostatically
  confined few-electron states}},\ }\href {https://doi.org/10.1063/5.0089350}
  {\bibfield  {journal} {\bibinfo  {journal} {AIP Adv.}\ }\textbf {\bibinfo
  {volume} {12}},\ \bibinfo {pages} {065123} (\bibinfo {year}
  {2022})}\BibitemShut {NoStop}%
\bibitem [{\citenamefont {Angus}\ \emph {et~al.}(2007)\citenamefont {Angus},
  \citenamefont {Ferguson}, \citenamefont {Dzurak},\ and\ \citenamefont
  {Clark}}]{angus2007gate}%
  \BibitemOpen
  \bibfield  {author} {\bibinfo {author} {\bibfnamefont {S.~J.}\ \bibnamefont
  {Angus}}, \bibinfo {author} {\bibfnamefont {A.~J.}\ \bibnamefont {Ferguson}},
  \bibinfo {author} {\bibfnamefont {A.~S.}\ \bibnamefont {Dzurak}},\ and\
  \bibinfo {author} {\bibfnamefont {R.~G.}\ \bibnamefont {Clark}},\ }\bibfield
  {title} {\bibinfo {title} {Gate-defined quantum dots in intrinsic silicon},\
  }\href {https://doi.org/10.1021/nl070949k} {\bibfield  {journal} {\bibinfo
  {journal} {Nano Lett.}\ }\textbf {\bibinfo {volume} {7}},\ \bibinfo {pages}
  {2051} (\bibinfo {year} {2007})}\BibitemShut {NoStop}%
\bibitem [{\citenamefont {Dodson}\ \emph {et~al.}(2020)\citenamefont {Dodson},
  \citenamefont {Holman}, \citenamefont {Thorgrimsson}, \citenamefont {Neyens},
  \citenamefont {MacQuarrie}, \citenamefont {McJunkin}, \citenamefont {Foote},
  \citenamefont {Edge}, \citenamefont {Coppersmith},\ and\ \citenamefont
  {Eriksson}}]{dodson2020fabrication}%
  \BibitemOpen
  \bibfield  {author} {\bibinfo {author} {\bibfnamefont {J.~P.}\ \bibnamefont
  {Dodson}}, \bibinfo {author} {\bibfnamefont {N.}~\bibnamefont {Holman}},
  \bibinfo {author} {\bibfnamefont {B.}~\bibnamefont {Thorgrimsson}}, \bibinfo
  {author} {\bibfnamefont {S.~F.}\ \bibnamefont {Neyens}}, \bibinfo {author}
  {\bibfnamefont {E.~R.}\ \bibnamefont {MacQuarrie}}, \bibinfo {author}
  {\bibfnamefont {T.}~\bibnamefont {McJunkin}}, \bibinfo {author}
  {\bibfnamefont {R.~H.}\ \bibnamefont {Foote}}, \bibinfo {author}
  {\bibfnamefont {L.~F.}\ \bibnamefont {Edge}}, \bibinfo {author}
  {\bibfnamefont {S.~N.}\ \bibnamefont {Coppersmith}},\ and\ \bibinfo {author}
  {\bibfnamefont {M.~A.}\ \bibnamefont {Eriksson}},\ }\bibfield  {title}
  {\bibinfo {title} {Fabrication process and failure analysis for robust
  quantum dots in silicon},\ }\href {https://doi.org/10.1088/1361-6528/abb559}
  {\bibfield  {journal} {\bibinfo  {journal} {Nanotechnology}\ }\textbf
  {\bibinfo {volume} {31}},\ \bibinfo {pages} {505001} (\bibinfo {year}
  {2020})}\BibitemShut {NoStop}%
\bibitem [{\citenamefont {Philips}\ \emph {et~al.}(2022)\citenamefont
  {Philips}, \citenamefont {Mądzik}, \citenamefont {Amitonov}, \citenamefont
  {de~Snoo}, \citenamefont {Russ}, \citenamefont {Kalhor}, \citenamefont
  {Volk}, \citenamefont {Lawrie}, \citenamefont {Brousse}, \citenamefont
  {Tryputen}, \citenamefont {Wuetz}, \citenamefont {Sammak}, \citenamefont
  {Veldhorst}, \citenamefont {Scappucci},\ and\ \citenamefont
  {Vandersypen}}]{philips2022universal}%
  \BibitemOpen
  \bibfield  {author} {\bibinfo {author} {\bibfnamefont {S.~G.~J.}\
  \bibnamefont {Philips}}, \bibinfo {author} {\bibfnamefont {M.~T.}\
  \bibnamefont {Mądzik}}, \bibinfo {author} {\bibfnamefont {S.~V.}\
  \bibnamefont {Amitonov}}, \bibinfo {author} {\bibfnamefont {S.~L.}\
  \bibnamefont {de~Snoo}}, \bibinfo {author} {\bibfnamefont {M.}~\bibnamefont
  {Russ}}, \bibinfo {author} {\bibfnamefont {N.}~\bibnamefont {Kalhor}},
  \bibinfo {author} {\bibfnamefont {C.}~\bibnamefont {Volk}}, \bibinfo {author}
  {\bibfnamefont {W.~I.~L.}\ \bibnamefont {Lawrie}}, \bibinfo {author}
  {\bibfnamefont {D.}~\bibnamefont {Brousse}}, \bibinfo {author} {\bibfnamefont
  {L.}~\bibnamefont {Tryputen}}, \bibinfo {author} {\bibfnamefont {B.~P.}\
  \bibnamefont {Wuetz}}, \bibinfo {author} {\bibfnamefont {A.}~\bibnamefont
  {Sammak}}, \bibinfo {author} {\bibfnamefont {M.}~\bibnamefont {Veldhorst}},
  \bibinfo {author} {\bibfnamefont {G.}~\bibnamefont {Scappucci}},\ and\
  \bibinfo {author} {\bibfnamefont {L.~M.~K.}\ \bibnamefont {Vandersypen}},\
  }\bibfield  {title} {\bibinfo {title} {Universal control of a six-qubit
  quantum processor in silicon},\ }\href
  {https://doi.org/10.1038/s41586-022-05117-x} {\bibfield  {journal} {\bibinfo
  {journal} {Nature}\ }\textbf {\bibinfo {volume} {609}},\ \bibinfo {pages}
  {919} (\bibinfo {year} {2022})}\BibitemShut {NoStop}%
\bibitem [{\citenamefont {Cai}\ \emph {et~al.}(2023)\citenamefont {Cai},
  \citenamefont {Connors}, \citenamefont {Edge},\ and\ \citenamefont
  {Nichol}}]{cai2023coherent}%
  \BibitemOpen
  \bibfield  {author} {\bibinfo {author} {\bibfnamefont {X.}~\bibnamefont
  {Cai}}, \bibinfo {author} {\bibfnamefont {E.~J.}\ \bibnamefont {Connors}},
  \bibinfo {author} {\bibfnamefont {L.~F.}\ \bibnamefont {Edge}},\ and\
  \bibinfo {author} {\bibfnamefont {J.~M.}\ \bibnamefont {Nichol}},\ }\bibfield
   {title} {\bibinfo {title} {Coherent spin–valley oscillations in silicon},\
  }\href {https://doi.org/10.1038/s41567-022-01870-y} {\bibfield  {journal}
  {\bibinfo  {journal} {Nat. Phys.}\ }\textbf {\bibinfo {volume} {19}},\
  \bibinfo {pages} {386} (\bibinfo {year} {2023})}\BibitemShut {NoStop}%
\bibitem [{\citenamefont {Anderson}(2009)}]{anderson2009efficient}%
  \BibitemOpen
  \bibfield  {author} {\bibinfo {author} {\bibfnamefont {C.~R.}\ \bibnamefont
  {Anderson}},\ }\bibfield  {title} {\bibinfo {title} {{Efficient solution of
  the Schroedinger–Poisson equations in layered semiconductor devices}},\
  }\href {https://doi.org/https://doi.org/10.1016/j.jcp.2009.03.037} {\bibfield
   {journal} {\bibinfo  {journal} {J. Comput. Phys.}\ }\textbf {\bibinfo
  {volume} {228}},\ \bibinfo {pages} {4745} (\bibinfo {year}
  {2009})}\BibitemShut {NoStop}%
\bibitem [{\citenamefont {Keyes}\ \emph {et~al.}(1995)\citenamefont {Keyes},
  \citenamefont {Saad},\ and\ \citenamefont {Truhlar}}]{keyes1995domain}%
  \BibitemOpen
  \bibfield  {author} {\bibinfo {author} {\bibfnamefont {D.~E.}\ \bibnamefont
  {Keyes}}, \bibinfo {author} {\bibfnamefont {Y.}~\bibnamefont {Saad}},\ and\
  \bibinfo {author} {\bibfnamefont {D.~G.}\ \bibnamefont {Truhlar}},\ }\href
  {https://doi.org/10.1137/1.9781611971507} {\emph {\bibinfo {title}
  {Domain-Based Parallelism and Problem Decomposition Methods in Computational
  Science and Engineering}}}\ (\bibinfo  {publisher} {SIAM},\ \bibinfo {year}
  {1995})\BibitemShut {NoStop}%
\bibitem [{\citenamefont {Boykin}\ \emph {et~al.}(2004)\citenamefont {Boykin},
  \citenamefont {Klimeck}, \citenamefont {Friesen}, \citenamefont
  {Coppersmith}, \citenamefont {von Allmen}, \citenamefont {Oyafuso},\ and\
  \citenamefont {Lee}}]{boykin2004valley}%
  \BibitemOpen
  \bibfield  {author} {\bibinfo {author} {\bibfnamefont {T.~B.}\ \bibnamefont
  {Boykin}}, \bibinfo {author} {\bibfnamefont {G.}~\bibnamefont {Klimeck}},
  \bibinfo {author} {\bibfnamefont {M.}~\bibnamefont {Friesen}}, \bibinfo
  {author} {\bibfnamefont {S.~N.}\ \bibnamefont {Coppersmith}}, \bibinfo
  {author} {\bibfnamefont {P.}~\bibnamefont {von Allmen}}, \bibinfo {author}
  {\bibfnamefont {F.}~\bibnamefont {Oyafuso}},\ and\ \bibinfo {author}
  {\bibfnamefont {S.}~\bibnamefont {Lee}},\ }\bibfield  {title} {\bibinfo
  {title} {Valley splitting in low-density quantum-confined heterostructures
  studied using tight-binding models},\ }\href
  {https://doi.org/10.1103/PhysRevB.70.165325} {\bibfield  {journal} {\bibinfo
  {journal} {Phys. Rev. B}\ }\textbf {\bibinfo {volume} {70}},\ \bibinfo
  {pages} {165325} (\bibinfo {year} {2004})}\BibitemShut {NoStop}%
\bibitem [{\citenamefont {Jiang}\ \emph {et~al.}(2012)\citenamefont {Jiang},
  \citenamefont {Kharche}, \citenamefont {Boykin},\ and\ \citenamefont
  {Klimeck}}]{Jiang2012effects}%
  \BibitemOpen
  \bibfield  {author} {\bibinfo {author} {\bibfnamefont {Z.}~\bibnamefont
  {Jiang}}, \bibinfo {author} {\bibfnamefont {N.}~\bibnamefont {Kharche}},
  \bibinfo {author} {\bibfnamefont {T.}~\bibnamefont {Boykin}},\ and\ \bibinfo
  {author} {\bibfnamefont {G.}~\bibnamefont {Klimeck}},\ }\bibfield  {title}
  {\bibinfo {title} {{Effects of interface disorder on valley splitting in
  SiGe/Si/SiGe quantum wells}},\ }\href {https://doi.org/10.1063/1.3692174}
  {\bibfield  {journal} {\bibinfo  {journal} {Appl. Phys. Lett.}\ }\textbf
  {\bibinfo {volume} {100}},\ \bibinfo {pages} {103502} (\bibinfo {year}
  {2012})}\BibitemShut {NoStop}%
\bibitem [{\citenamefont {Abadillo-Uriel}\ \emph {et~al.}(2018)\citenamefont
  {Abadillo-Uriel}, \citenamefont {Thorgrimsson}, \citenamefont {Kim},
  \citenamefont {Smith}, \citenamefont {Simmons}, \citenamefont {Ward},
  \citenamefont {Foote}, \citenamefont {Corrigan}, \citenamefont {Savage},
  \citenamefont {Lagally}, \citenamefont {Calder\'on}, \citenamefont
  {Coppersmith}, \citenamefont {Eriksson},\ and\ \citenamefont
  {Friesen}}]{Abadillo-Uriel2018signatures}%
  \BibitemOpen
  \bibfield  {author} {\bibinfo {author} {\bibfnamefont {J.~C.}\ \bibnamefont
  {Abadillo-Uriel}}, \bibinfo {author} {\bibfnamefont {B.}~\bibnamefont
  {Thorgrimsson}}, \bibinfo {author} {\bibfnamefont {D.}~\bibnamefont {Kim}},
  \bibinfo {author} {\bibfnamefont {L.~W.}\ \bibnamefont {Smith}}, \bibinfo
  {author} {\bibfnamefont {C.~B.}\ \bibnamefont {Simmons}}, \bibinfo {author}
  {\bibfnamefont {D.~R.}\ \bibnamefont {Ward}}, \bibinfo {author}
  {\bibfnamefont {R.~H.}\ \bibnamefont {Foote}}, \bibinfo {author}
  {\bibfnamefont {J.}~\bibnamefont {Corrigan}}, \bibinfo {author}
  {\bibfnamefont {D.~E.}\ \bibnamefont {Savage}}, \bibinfo {author}
  {\bibfnamefont {M.~G.}\ \bibnamefont {Lagally}}, \bibinfo {author}
  {\bibfnamefont {M.~J.}\ \bibnamefont {Calder\'on}}, \bibinfo {author}
  {\bibfnamefont {S.~N.}\ \bibnamefont {Coppersmith}}, \bibinfo {author}
  {\bibfnamefont {M.~A.}\ \bibnamefont {Eriksson}},\ and\ \bibinfo {author}
  {\bibfnamefont {M.}~\bibnamefont {Friesen}},\ }\bibfield  {title} {\bibinfo
  {title} {{Signatures of atomic-scale structure in the energy dispersion and
  coherence of a Si quantum-dot qubit}},\ }\href
  {https://doi.org/10.1103/PhysRevB.98.165438} {\bibfield  {journal} {\bibinfo
  {journal} {Phys. Rev. B}\ }\textbf {\bibinfo {volume} {98}},\ \bibinfo
  {pages} {165438} (\bibinfo {year} {2018})}\BibitemShut {NoStop}%
\end{thebibliography}%

\end{document}